\documentclass[12pt]{article}
\usepackage{amsmath}
\usepackage{amsfonts}
\usepackage{amssymb}
\usepackage{graphicx}
\providecommand{\U}[1]{\protect\rule{.1in}{.1in}}
\providecommand{\U}[1]{\protect\rule{.1in}{.1in}}
\providecommand{\U}[1]{\protect\rule{.1in}{.1in}}
\providecommand{\U}[1]{\protect\rule{.1in}{.1in}}
\providecommand{\U}[1]{\protect\rule{.1in}{.1in}}
\providecommand{\U}[1]{\protect\rule{.1in}{.1in}}
\providecommand{\U}[1]{\protect\rule{.1in}{.1in}}
\providecommand{\U}[1]{\protect\rule{.1in}{.1in}} \textwidth17.1cm \oddsidemargin -0.5cm

\newcommand{\be}{\begin{equation}}
\newcommand{\ee}{\end{equation}}
\newcommand{\ben}{\begin{equation*}}
\newcommand{\een}{\end{equation*}}
\newcommand{\ar}{\begin{array}}
\newcommand{\arn}{\end{array}}

\newcommand{\vk}{\vec{k}}
\newcommand{\vks}{\vec{k}^{\;2}}
\newcommand{\q}{\vec{q}}
\newcommand{\qs}{\vec{q}^{\;2}}
\newcommand{\qp}{\vec{q}^{\;\prime}}
\newcommand{\qps}{\vec{q}^{\;\prime\; 2}}
\newcommand{\x}{\vec{r}}

\newcommand{\xp}{\vec{r}^{\;\prime}}

\newcommand{\vrho}{\vec\rho}

\def\pnot{\mbox{${\not{\hbox{\kern-3.0pt$p$}}}$}}
\def\qnot{\mbox{${\not{\hbox{\kern-2.0pt$q$}}}$}}
\def\enot{\mbox{${\not{\hbox{\kern-2.0pt$e$}}}$}}
\def\knot{\mbox{${\not{\hbox{\kern-2.0pt$k$}}}$}}

\def\fun#1#2{\lower3.6pt\vbox{\baselineskip0pt\lineskip.9pt\ialign
{$\mathsurround=0pt#1\hfil##\hfil$\crcr#2\crcr\sim\crcr}}}

\begin{document}

\begin{titlepage}

\begin{center}
{\bf The dipole form of the  BFKL kernel in supersymmetric
Yang--Mills theories$^{~\ast}$}
\end{center}
\vskip 0.5cm \centerline{V.S.~Fadin$^{a\,\dag}$,
R.~Fiore$^{b\,\ddag}$} \vskip .6cm

\centerline{\sl $^{a}$ Budker Institute of Nuclear Physics, 630090
Novosibirsk, Russia} \centerline{\sl Novosibirsk State University,
630090 Novosibirsk, Russia} \centerline{\sl $^{b}$ Dipartimento di
Fisica, Universit\`a della Calabria,} \centerline{\sl Istituto
Nazionale di Fisica Nucleare, Gruppo collegato di Cosenza,}
\centerline{\sl Arcavacata di Rende, I-87036 Cosenza, Italy}
\vskip 2cm

\begin{abstract}
The dipole (M\"{o}bius) representation of the colour singlet BFKL
kernel in the next-to-leading order is found in supersymmetric
Yang--Mills theories. Ambiguities of this form and its conformal
properties are discussed.

\end{abstract}
\vfill \hrule \vskip.3cm \noindent $^{\ast}${\it Work supported in
part by the RFBR grant 07-02-00953, in part by the RFBR--MSTI
grant 06-02-72041, in part by INTAS and in part by Ministero
Italiano dell'Universit\`a e della Ricerca.} \vfill $
\begin{array}{ll} ^{\dag}\mbox{{\it e-mail address:}} &
\mbox{FADIN@INP.NSK.SU}\\
^{\ddag}\mbox{{\it e-mail address:}} &
\mbox{FIORE@CS.INFN.IT}\\
\end{array}
$

\end{titlepage}

\vfill \eject

\section{Introduction}

For scattering of calourless objects the kernel $\hat{\cal K}$ of
the BFKL~\cite{BFKL} equation can be taken in a special form,
which we call dipole form~\cite{Fadin:2006ha}.  This form is
obtained by transformation of the kernel $\langle
\q_1\q_2|\hat{{\cal K}}|\qp_1\qp_2\rangle$ from the space of
transverse momenta $\q_i, \;\;i=1,2$, where it is originally
defined, to the space of transverse coordinates $\x_i$, with
subsequent  rejection of the  terms proportional to
$\delta(\xp_1-\xp_2)$ in $\langle \x_1\x_2|\hat{{\cal
K}}|\xp_1\xp_2\rangle$ and revising terms not depending either on
$\x_1$ or on  $\x_2$ so as to make the kernel vanishing at
$\x_1=\x_2$ (see Ref.~\cite{Fadin:2006ha} for details). In another
words, the kernel can be considered as acting in the space of
functions, vanishing at $\x_1=\x_2$. This space is called
M\"{o}bius space~\cite{Bartels:2004ef} and the dipole form is
called also M\"{o}bius form. We will use both names and denote
this form as $\langle \x_1\x_2|\hat{{\cal
K}}_d|\xp_1\xp_2\rangle$.

The M\"{o}bius form is interesting for several reasons. One of
them is the investigation of the inter-relation of the BFKL
approach and the colour dipole model~\cite{dipole}, where  the
nonlinear generalization of the BFKL equation ~\cite{Balitsky} is
obtained. A clear undestanding of this inter-relation  can be
helpful for the further development of the theoretical description
of small-x processes. Another reason is the study of conformal
properties of the kernel. It is known~\cite{Lipatov:1985uk} that
in the leading order (LO) the BFKL kernel in the M\"{o}bius space
is conformal invariant, property which is extremely useful for
finding solutions of the BFKL equation. Evidently, in the
next-to-leading order (NLO) in QCD conformal invariance is
violated by renormalization. However, one can expect conformal
invariance of the NLO BFKL kernel in supersymmetric extensions of
QCD. An additional reason is a possible simplification of the
kernel in the  M\"{o}bius form compared to the  momentum space
representation.

In Refs.~\cite{Fadin:2006ha},~\cite{Fadin:2007ee}
and~\cite{Fadin:2007de} the M\"{o}bius form of the quark and gluon
parts of the kernel correspondingly was found for the QCD case. It
was shown that the dipole form of the quark part agrees with the
result obtained in Ref.~\cite{Balitsky:2006wa} by the direct
calculation of the corresponding contribution to the
Balitsky-Kovchegov (BK) kernel~\cite{Balitsky}. It was shown also
that the M\"{o}bius form of the ``Abelian" piece of the quark
part, which is not associated with renormalization and has the
most complicated form in the momentum space, is strongly
simplified and is  conformal invariant. As well as the quark part,
the gluon part in the dipole form is greatly simplified compared
with the momentum space representation, although conformal
invariance of this part is broken by several terms, among which
those not associated  with renormalization. However, the ambiguity
of the NLO kernel~\cite{Fadin:2006ha} (analogous to the ambiguity
of the NLO anomalous dimensions), allowing the transformations
\be\hat{\cal K}\rightarrow \hat{\cal K} -[\hat{\cal
K}^{B}\hat{O}]~,\label{trans}\ee where $\hat{\cal K}^{B}$  is the
LO kernel and $\hat{O}\sim g^{2}$, leaves a hope for conformal
invariance of the piece not related to  renormalization.

The aim of this work is to consider supersymmetric Yang--Mills
theories. We analyze the generalizations of the BFKL kernel for
these theories and  find their  dipole forms.

\section{An overall view of the M\"{o}bius form}

We use the same notation as in Ref.~\cite{Fadin:2006ha}: $\vec{q}
_{i}^{\;\prime}$ and $\vec{q}_{i}$, $i=1,2$, represent the
transverse momenta of Reggeons in initial and final $t$-channel
states, while $\vec{r} _{i}^{\;\prime}$ and $\vec{r}_{i}$ are the
corresponding conjugate coordinates. The state normalization is
\begin{equation}
\langle\vec{q}|\vec{q}^{\;\prime}\rangle=\delta(\vec{q}-\vec{q}^{\;\prime
})\;,\;\;\;\;\;\langle\vec{r}|\vec{r}^{\;\prime}\rangle=\delta(\vec{r}-\vec
{r}^{\;\prime})\;, \label{normalization}
\end{equation}
so that
\begin{equation}
\langle\vec{r}|\vec{q}\rangle=\frac{e^{i\vec{q}\,\vec{r}}}{(2\pi)^{1+\epsilon
}}\;,
\end{equation}
where $\epsilon=(D-4)/2$; $D-2$ is the dimension of the transverse
space and it is taken different from $2$ for the regularization of
divergences. We shall also use the notation
$\vec{q}=\vec{q}_{1}+\vec{q}_{2},\;\;\vec{q}^{\;\prime
}=\vec{q}_{1}^{\;\prime}+\vec{q}_{2}^{\;\prime};\;\;\vec{k}=\vec{q}_{1}
-\vec{q}_{1}^{\;\prime}=\vec{q}_{2}^{\;\prime}-\vec{q}_{2}$.

In the LO, discussed in detail in Ref.~\cite{Fadin:2006ha}, the
dipole form is written as
\begin{equation}
\langle\vec{r}_{1}\vec{r}_{2}|\hat{\mathcal{K}}_{d}^{LO}|\vec{r}_{1}
^{\;\prime}\vec{r}_{2}^{\;\prime}\rangle=\frac{\alpha_{s}(\mu)N_{c}}{2\pi^{2}
}\int
d\vec{\rho}\frac{\vec{r}_{12}{}^{2}}{\vec{r}_{1\rho}^{\,\,2}\vec
{r}_{2\rho}^{\,\,2}}\Biggl[\delta(\vec{r}_{11^{\prime}})\delta(\vec
{r}_{2^{\prime}\rho})+\delta(\vec{r}_{1^{\prime}\rho})\delta(\vec
{r}_{22^{\prime}})-\delta(\vec{r}_{11^{\prime}})\delta({r}_{22^{\prime}
})\Biggr]~. \label{LO dipole}
\end{equation}
Here and below  $\x_{ij}=\x_{i}-\x_{j},\;\;\x_{i^\prime
j^\prime}=\x_{i}^{~\prime}-\x_{j}^{~\prime}, \;\;
\x_{ij^\prime}=\x_{i}-\x_{j}^{\prime}, \;\;\x_{i\rho}=\x_{i}-\vec
\rho$. Note that the integrand in Eq.~(\ref{LO dipole}) contains
ultraviolet singularities at $\vec{\rho}=\vec{r}_{1}$ and
$\vec{\rho}=\vec{r}_{2}$ which cancel in the sum of the
contributions assuming that the kernel acts in the M\"{o}bius
space. The coefficient of $\delta(\vec
{r}_{11^{\prime}})\delta({\vec{r}}_{22^{\prime}})$ is written in
the integral form in order to make the cancellation evident. The
singularities do not permit us to perform the integration in this
coefficient.  In Eq.~(\ref{LO dipole}) $\alpha_s(\mu)=g_\mu^2/(4\pi)$,
where $\; g_\mu$ is the renormalized coupling and the argument $\mu$ is
shown because we want to present the NLO corrections, which depend
on this argument.

A general view of the NLO dipole form is
\[
\langle\vec{r}_{1}\vec{r}_{2}|\hat{\mathcal{K}}_{d}^{NLO}|\vec{r}
_{1}^{\;\prime}\vec{r}_{2}^{\;\prime}\rangle=\frac{\alpha_{s}^{2}(\mu
)N_{c}^{2}}{4\pi^{3}}\Biggl[\delta(\vec{r}_{11^{\prime}})\delta(\vec
{r}_{22^{\prime}})\int
d\vec{\rho}\,g^{0}(\vec{r}_{1},\vec{r}_{2};\vec{\rho})
\]
\begin{equation}
+\delta(\vec{r}_{11^{\prime}})g(\vec{r}_{1},\vec{r}_{2};\vec{r}_{2}^{\;\prime
})+\delta(\vec{r}_{22^{\prime}})g(\vec{r}_{2},\vec{r}_{1};\vec{r}
_{1}^{\;\prime})+\frac{1}{\pi}g(\vec{r}_{1},\vec{r}_{2};\vec{r}_{1}^{\;\prime
},\vec{r}_{2}^{\;\prime})\Biggr]\; \label{kernel through g}\;,
\end{equation}
with the functions $g$ turning into zero  when their first two
arguments coincide.  As well as in the LO, the coefficient of
$\delta(\vec{r}_{11^{\prime}})\delta({\vec{r}}_{22^{\prime}})$ is
written in the integral form in order to make evident the
cancellation of the ultraviolet divergencies. Of course, the
integrand in this form is not unique.  We shall use  the equalities
\begin{equation}
\int \frac{d\vec{\rho}}{\vec
{r}_{1\rho}^{\;2}\vec{r}_{2\rho}^{\;2}} \left[\vec{r}_{12}^{\;2}
\ln\left(\frac{\vec {r}_{1\rho}^{\;2}\vec{r}_{2\rho}^{\;2}}
{\vec{r}_{12}^{\;4}}\right) +\biggl(\vec
{r}_{1\rho}^{\;2}-\vec{r}_{2\rho}^{\;2}\biggr) \ln\left(\frac{\vec
{r}_{1\rho}^{\;2}}{\vec{r}_{2\rho}^{\;2}}\right)\right]=0\;,
\end{equation}
\begin{equation}
\frac{1}{4\pi}\int d\vec{\rho}\frac{\,\vec{r}_{12}^{\;2}}{\vec
{r}_{1\rho}^{\;2}\vec{r}_{2\rho}^{\;2}}\ln\left(
\frac{\vec{r}_{1\rho}^{\;2} }{\vec{r}_{12}^{\;2}}\right) \ln\left(
\frac{\vec{r}_{2\rho}^{\;2}}{\vec
{r}_{12}^{\;2}}\right)=\frac{1}{4\pi}\int
d\vec{\rho}\frac{1}{\vec{r}_{2\rho}^{\;2}}\ln\left(
\frac{\vec{r}_{1\rho}^{\;2} }{\vec{r}_{2\rho}^{\;2}}\right)
\ln\left( \frac{\vec{r}_{1\rho}^{\;2}}{\vec
{r}_{12}^{\;2}}\right)~, \label{useful relations for g 0}
\end{equation}
for modifying this integrand.

It is known (see Ref.~\cite{Brink:1976bc} and references therein)  that
in four dimensions only three types of  sypersymmetric Yang-Mills
theories ($N=1,$ $N=2$ and $N=4$ theories) are possible.  The
simplest, $N=1$ theory, contains the Yang-Mills fields (gluons)
and the Maiorana spinors (gluinos). At $N=2$ and $N=4$ besides the
vectors and spinors there are also scalars and pseudoscalars. For
our purposes there is no difference between scalars and
pseudoscalars. In the following we shall call both of them scalars.
All particles are contained in the adjoint representation of the
colour group. The number of gluinos $n_M$ is equal to $N$, the number
of scalars $n_S$ is equal to $2(N-1)$. We shall use the relation
between the bare coupling $g$ and the renormalized one $g_{\mu}$
in the ${\overline{\mbox{MS}}}$ form
\begin{equation}
g=g_{\mu}\mu^{-\mbox{\normalsize $\epsilon$}}\left[ 1+\frac
{{g}_{\mu}^{2}N_{c}}{(4\pi)^{2}}\left(\frac{11}{3}-\frac{2}{3}{n_M}-
\frac{1}{6}{n_S}\right)\frac{1}{2\epsilon}\right]~,
\label{coupling renormalization}
\end{equation}
where $n_M=N,\;\; n_S=2(N-1)$. It is known that the standard form
of the dimensional regularization violates supersymmetry, because
it violates the equality of Bose and Fermi degrees of freedom. The
modification of the dimensional regularization which preserves
sypersymmetry (it is called dimensional reduction) was suggested
in Ref.~\cite{Siegel-79}. Unfortunately, it contains the inherent
inconsistency~\cite{Siegel-80}.  However, the inconsistency
becomes apparent only at the three-loop level and can be removed
with the violation of supersymmetry in the highest
orders~\cite{Avdeev:1982xy}. Therefore the dimensional reduction
is widely used for practical calculations. With our accuracy the
dimensional reduction is equivalent to the  dimensional
regularization with the number of scalars $n_S$ equal to
$2(N-1)-2\epsilon$. It follows from Eq.~(\ref{coupling
renormalization}) that the use of the dimensional reduction
instead of the dimensional regularization is equivalent to the
finite charge remornalizaion \be \alpha_s(\mu)\rightarrow
\alpha_s(\mu)\left(1-\frac{\alpha_s(\mu)N_c}{12\pi}\right).\label{finite
renormalization}\ee

In Refs.~\cite{Fadin:2006ha} and \cite{Fadin:2007de} the operator
transformation (\ref{trans}), with the operator $\hat{O}$
associated with the charge renormalization, was applied to the
kernel in the momentum space defined by the prescriptions given in
Ref.~\cite{FF98}. It occurs~\cite{Fadin:2006ha} that this
transformation simplifies considerably the quark part.  We shall
use the SUSY generalization of this transformation, with \be
\hat{O}=\hat{O}_G+\hat{O}_M+\hat{O}_S=-\frac{\alpha_{s}N_c}{8\pi}
\left(\frac{11}{3}-\frac{2}{3}n_M-\frac{1}{6}n_S\right)
\ln\left(\frac{\hat{\vec{q}}_{1}^{\;2}\hat{\vec{q}}_{2}^{\;2}}
{\mu^4}\right). \label{trans-1}\ee At the $N=4$ supersymmetry
${11}/{3}-({2}/{3})n_M- ({1}/{6})n_S=0, $ $\alpha_s$ does not
depend on $\mu$  and  $\hat{O}=0$.

\section{Gluon and gluino parts}
The gluon contribution to the kernel is  the same as in QCD. We
denote this part by the subscript $G$. Using Eq.~(\ref{useful
relations for g 0}), from the results of Ref.~\cite{Fadin:2007de}
we obtain
\be
g^0_G(\vec{r}_{1},\vec{r}_{2};\vec{\rho})\
=-g_G(\vec{r}_{1},\vec{r}_{2};\vec{\rho})\
+\frac{1}{2}\frac{\,\vec{r}_{12}^{\;2}}{\vec
{r}_{1\rho}^{\;2}\vec{r}_{2\rho}^{\;2}}\ln\left(
\frac{\vec{r}_{1\rho}^{\;2} }{\vec{r}_{12}^{\;2}}\right) \ln\left(
\frac{\vec{r}_{2\rho}^{\;2}}{\vec {r}_{12}^{\;2}}\right)~,\label{g
0 G} \ee
\[
g_G(\vec{r}_{1},\vec{r}_{2};\vec{\rho})\ =\frac{11}{6}\frac{\vec
{r}_{12}^{\;2}}{\vec{r}_{2\rho}^{\;2}\vec{r}_{1\rho}^{\;2}}
\ln\left( \frac{\vec{r}_{12}^{\;2}}{r_{G}^{2}}\right)
+\frac{11}{6}\frac{\vec{r}_{1\rho
}^{\,\,\,2}-\vec{r}_{2\rho}^{\,\,\,2}}{\vec{r}_{1\rho
}^{\,\,\,2}\vec{r}_{2\rho}^{\,\,\,2}} \ln\left(
\frac{\vec{r}_{2\rho}^{\,\,\,2}}{\vec
{r}_{1\rho}^{\,\,\,2}}\right)
\]
\be +\frac{1}{2\vec{r}_{2\rho}^{\;2}}\ln\left(
\frac{\vec{r}_{1\rho}^{\;2}}{\vec{r}_{2\rho}^{\;2}}\right)
\ln\left( \frac{\vec{r} _{12}^{\;2}}{\vec{r}_{1\rho}^{\;2}}\right)
-\frac{\vec{r}_{12}^{\;2}
}{2\,\vec{r}_{2\rho}^{\;2}\vec{r}_{1\rho}^{\;2}}\ln\left(
\frac{\vec{r}_{12}^{\;2}}{\vec{r}_{2\rho}^{\;2}}\right) \ln\left(
\frac{\vec{r}_{12}^{\;2}}{\vec{r}_{1\rho}^{\;2}}\right)  , \ee
where \be \ln
r_{G}^{2}=2\psi(1)-\ln\frac{\mu^{2}}{4}-\frac{3}{11}\left(
\frac{67} {9}-2\zeta(2)\right)  .\ee Both
$g_G^{0}(\vec{r}_{1},\vec{r}_{2};\vec{\rho})$ and
$g_G(\vec{r}_{1},\vec{r}_{2};\vec{\rho})$ {vanish at
$\vec{r}_{1}=\vec{r}_{2}$}. Then, these functions {turn into zero
for $\vec{\rho}^{\;2}\rightarrow\infty$} faster than
$(\vec{\rho}^{\;2})^{-1}$ to provide the infrared safety. The
{ultraviolet singularities} of these functions at
$\vec{\rho}=\vec{r}_{2}$\ and $\vec{\rho}=\vec{r}_{1}$ cancel
assuming that the kernel acts in the M\"{o}bius space.

The most complicated contribution is
$g_G(\vec{r}_{1},\vec{r}_{2};\vec{r}_{1}^{\;\prime},\vec{r}_{2}^{\;\prime
})$. It can be written as
\[
g_G(\vec{r}_{1},\vec{r}_{2};\vec{r}_{1}^{\;\prime
},\vec{r}_{2}^{\;\prime })= \frac{1}{\vec{r}_{1^{\prime }2^{\prime
}}^{\,\,4}}\left( \frac{\vec{r} _{11^{\prime
}}^{\;2}\,\vec{r}_{22^{\prime }}^{\;2}}{d}\ln \left( \frac{\vec{
r}_{12^{\prime }}^{\;2}\,\vec{r}_{21^{\prime
}}^{\;2}}{\vec{r}_{11^{\prime }}^{\;2}\vec{r}_{22^{\prime
}}^{\;2}}\right) -1\right)
+\frac{\vec{r}\,\,_{12}^{2}}{2d\,\vec{r}\,\,_{1^{\prime }2^{\prime
}}^{2}}\ln \left( \frac{\vec{r}\,\,_{12^{\prime
}}^{2}\vec{r}\,\,_{21^{\prime }}^{2}}{ \vec{r}\,\,_{11^{\prime
}}^{2}\ \vec{r}\,\,_{22^{\prime }}^{2}}\right) \left(
\frac{\vec{r}\,\,_{12}^{2}\vec{r}\,\,_{1^{\prime }2^{\prime
}}^{2}}{\vec{r}\,\,_{11^{\prime }}^{2}\ \vec{r}\,\,_{22^{\prime
}}^{2}}-4\right)
\]
\[
+\frac{1}{2\vec{r}\,\,_{12^{\prime }}^{2}\vec{r}\,\,_{21^{\prime
}}^{2}\ \vec{r}\,\,_{1^{\prime }2^{\prime
}}^{2}}\left(2(\vec{r}\,\,_{12^{\prime}}\vec{r}\,\,_{21^{\prime
}})\ln \left( \frac{\vec{r} \,\,_{11^{\prime
}}^{2}\vec{r}\,\,_{22^{\prime }}^{2}}{\vec{r}\,\,_{12}^{2}\
\vec{r}\,\,_{1^{\prime }2^{\prime }}^{2}}\right) +
\vec{r}\,\,_{12^{\prime }}^{2}\ln \left(
\frac{\vec{r}\,\,_{12}^{2}\vec{r}\,\,_{22^{\prime }}^{2}}{\
\vec{r}\,\,_{1^{\prime }2^{\prime }}^{2}\vec{r} \,\,_{11^{\prime
}}^{2}}\right)+ \vec{r}\,\,_{21^{\prime }}^{2}\ln \left(
\frac{\vec{r}\,\,_{12}^{2}\vec{r} \,\,_{11^{\prime }}^{2}}{\
\vec{r}\,\,_{1^{\prime }2^{\prime }}^{2}\vec{r}\,\,_{22^{\prime
}}^{2}}\right)\right)
\]
\[
+\frac{1}{2\vec{r}\,\,_{11^{\prime }}^{2}\vec{r}\,\,_{22^{\prime
}}^{2}\ \vec{r}\,\,_{1^{\prime }2^{\prime
}}^{2}}\left((\vec{r}\,\,_{11^{\prime
}}^{2}+\vec{r}\,\,_{22^{\prime }}^{2}+ \vec{r}\,\,_{12}^{2})\ln
\left( \frac{\vec{r}\,\,_{12}^{2}\ \vec{r}\,\,_{1^{\prime
}2^{\prime }}^{2}}{\vec{r} \,\,_{12^{\prime
}}^{2}\vec{r}\,\,_{21^{\prime }}^{2}}\right) +
(\vec{r}\,\,_{12^{\prime }}^{2}+\vec{r}\,\,_{21^{\prime }}^{2}-
\vec{r}\,\,_{1^{\prime }2^{\prime }}^{2}) \ln \left(
\frac{\vec{r}\,\,_{1^{\prime }2^{\prime
}}^{2}}{\vec{r}\,\,_{12}^{2}}\right)\right.
\]
\[
\left. + (\vec{r}\,\,_{12}^{2}- \vec{r}\,\,_{11^{\prime }}^{2})\ln
\left( \frac{\vec{r}\,\,_{11^{\prime }}^{2}}{\vec{r}
\,\,_{12^{\prime }}^{2}}\right)+ (\vec{r}\,\,_{12}^{2}-
\vec{r}\,\,_{22^{\prime }}^{2})\ln \left(
\frac{\vec{r}\,\,_{22^{\prime }}^{2}}{\vec{r} \,\,_{21^{\prime
}}^{2}}\right)+ \vec{r}\,\,_{21^{\prime }}^{2}\ln \left(
\frac{\vec{r}\,\,_{21^{\prime }}^{2}}{\vec{r} \,\,_{11^{\prime
}}^{2}}\right)+ \vec{r}\,\,_{12^{\prime }}^{2}\ln \left(
\frac{\vec{r}\,\,_{12^{\prime }}^{2}}{\vec{r} \,\,_{22^{\prime
}}^{2}}\right) \right)
\]
\begin{equation}
+\frac{1}{2\vec{r}\,\,_{11^{\prime }}^{2}\vec{r}\,\,_{22^{\prime
}}^{2}}\left(\frac{\vec{r}\,\,_{12}^{2}}{\vec{r}\,\,_{12^{\prime
}}^{2}}\ln \left( \frac{\vec{r}\,\,_{11^{\prime }}^{2}}{\vec{r}
\,\,_{1^{\prime }2^{\prime
}}^{2}}\right)+\frac{\vec{r}\,\,_{12}^{2}}{\vec{r}\,\,_{21^{\prime
}}^{2}}\ln \left( \frac{\vec{r}\,\,_{22^{\prime }}^{2}}{\vec{r}
\,\,_{1^{\prime }2^{\prime
}}^{2}}\right)+\frac{\vec{r}\,\,_{22^{\prime
}}^{2}}{\vec{r}\,\,_{12^{\prime }}^{2}}\ln \left(
\frac{\vec{r}\,\,_{22^{\prime }}^{2}}{\vec{r}
\,\,_{12}^{2}}\right)+\frac{\vec{r}\,\,_{11^{\prime
}}^{2}}{\vec{r}\,\,_{21^{\prime }}^{2}}\ln \left(
\frac{\vec{r}\,\,_{11^{\prime }}^{2}}{\vec{r}
\,\,_{12}^{2}}\right) \right),
\end{equation}
where \be
d=\vec{r}_{12^{\prime }}^{\;2}\vec{r}_{21^{\prime }}^{\;2}-\vec{r}%
_{11^{\prime }}^{\;2}\vec{r}_{22^{\prime }}^{\,\,2}~.\label{d} \ee
The gluino part can be obtained from the results of
Refs.~\cite{Fadin:2006ha} and \cite{Fadin:2007ee} by the change of
the coefficients: $ n_f\rightarrow n_M N_c\;$ for the
``non-Abelian" part and $n_f\rightarrow -n_M N^3_c \;$ for the
``Abelian" part. Using the subscript $M$ to denote this part, we
have  \be g_M(\x_1,\x_2;\vrho)=-g^0_M(\x_1,\x_2;\vrho)=
\frac{n_M}{3}\left(\frac{\x_{12}^2}{\x_{1\rho}^2\x_{2\rho}^2}\ln\frac{\x_{M}^{\,2}}
{\x_{12}^2}+\frac{\x_{1\rho}^2-\x_{2\rho}^2}{\x_{1\rho}^2\x_{2\rho}^2}
\ln\frac{\x_{1\rho}^2}{\x_{2\rho}^2}\right)~, \ee where
\be\ln\x_{M}^{\,2}=-\frac{5}{3}+2\psi(1)-\ln\frac{\mu^2}{4} \ee
and \be
g_{M}(\x_1,\x_2;\xp_1,\xp_2)=-{n_M}\frac{1}{\x^{\,4}_{1'2'}}
\biggl(\frac{\x^{\,2}_{12'}\x^{\,2}_{1'2}+\x^{\,2}_{11'}\x^{\,2}_{22'}
-\x^{\,2}_{12}\x^{\,2}_{1'2'}}{2d}\ln\frac{\x^{\,2}_{12'}
\x^{\,2}_{1'2}}{\x^{\,2}_{11'}\x^{\,2}_{22'}}-1\biggr)~. \label{g4
M} \ee

\section{Dipole form of the scalar part}

\label{sec:scalar momentum}

Let us present the scalar contribution to the kernel in the
momentum space. Details of the derivation will be given
elsewhere~\cite{FG}.

The scalar particles contribute both to the gluon Regge trajectory
$\omega(\q)$ and to the ``real" part $\hat{{\cal K}}_r$ of the
kernel: \be \langle \x_1\x_2|\hat{{\cal K}}^{S}|\xp_1\xp_2\rangle
=
\delta(\q_1-\qp_1)\delta(\q_2-\qp_2)\left(\omega_S(\q_1)+\omega_S(\q_2)\right)+\langle
\x_1\x_2|\hat{{\cal K}}_r^{S}|\xp_1\xp_2\rangle~. \ee The
contribution to the trajectory can be obtained from the quark
contribution through the substitution $n_f\rightarrow {n_s}{N_c}$
and the change of the fermion polarization operator with the
scalar one:
\begin{equation}
\omega_S(\q)=\frac{g^4\qs N^2_cn_S}{4(2\pi )^{D-1}}\int
\frac{d^{(D-2)}k}{\vec k^{\:2}(\vec k-\vec q \,)^{\:2}}\left[
P_s(\vec q \,)-P_s(\vec k)-P_s(\q-\vec k) \right] ~, \label{omega
s}
\end{equation}
where
\begin{equation}
P_s(\vec q)=\frac{2\Gamma \left(1-\epsilon\right) \Gamma
^2\left(2+\epsilon\right)}{(4\pi
)^{2+\epsilon}\epsilon(1+\epsilon)\Gamma \left(4+\epsilon\right)}
\vec q^{\:2\epsilon}~. \label{polarization s}
\end{equation}
Here $P_s(\vec q)$ is  the scalar polarization operator where the
coupling constant and colour/flavour coefficients have been omitted. It
differs from the corresponding fermion operator  by the factor
$1/(4(1+\epsilon))$.

It is convenient to divide the scalar contribution to the ``real''
kernel into the ``non-Abelian" and ``Abelian" parts, $\hat{\cal{
K}}^S_r=\hat{\cal{ K}}^S_n+\hat{\cal{ K}}^S_a$, quite analogously
to the quark case~\cite{FFP99}. Besides the contribution from
production of a couple of scalars,  the ``non-Abelian" part
contains also the scalar loop contribution to one-gluon
production. Separately these contributions are rather complicated,
but taken together they acquire  a simple form:
\[
\langle\q_{1}\q_{2}|\hat{\cal{ K}}^S_n|\qp_1\qp_2 \rangle=
=\delta(\q_{1}+\q_{2}-\qp_{1}-\qp_{2})\frac{g^{4}N_c^2n_S}{4(2\pi
)^{D-1}}\left[2\left(\frac{\qps_1}{\qs_1\vks}+\frac{\qps_2}{\qs_2\vks}-
\frac{\qs}{\qs_1\qs_2}\right)P_s(\vk)\right.
\]
\[
\left.+\frac{\qs}{\qs_1\qs_2}\left(2P_s(\vk)+2P_s(\q)-P_s(\q_1)-P_s(\q_2)-P_s(\qp_1)
-P_s(\qp_2)\right) \right.
\]
\begin{equation}
\left.
+\left(\frac{\qps_1}{\qs_1\vks}-\frac{\qps_2}{\qs_2\vks}\right)
\left(P_s(\q_1)-P_s(\q_2)-P_s(\qp_1)+P_s(\qp_2)\right)\right]~.
\label{K s n-A}
\end{equation}
In accordance with the results of Ref.~\cite{Kotikov:2000pm}, it
can be obtained from the ``non-Abelian" quark
contribution~\cite{FFP99} through the same substitutions as in the
case of the trajectory.  Note that the result (\ref{K s n-A}) is
obtained with the prescriptions given in Ref.~\cite{FF98}. As it
was explained, we shall apply to it the transformation
(\ref{trans}) with the operator $\hat{O}$ given by the formula
(\ref{trans-1}).

In Eqs.~(\ref{omega s}) and (\ref{K s n-A}) the space-time
dimension $D$ is taken different from $4$ to regularize the
infrared and ultraviolet divergencies. The last ones are removed
by the renormalization (\ref{coupling renormalization}) of the
coupling constant both in the leading order trajectory and ``real"
Born kernel: \be {\omega(\q)} =-\frac{g^2N_c \vec
q^{\:2}}{2(2\pi)^{D-1}} \int\frac{d^{D-2}k}{\vec k^{\:2}(\vec
q-\vec k)^{2}}~,\;\; \label{omega B} \ee \be \langle
\q_{1},\q_{2}|\hat{\cal K}^B_r|
\qp_{1},\qp_{2}\rangle=\delta(\q-\qp)\frac{1}{\qs_1\qs_2}\frac{g^2N_c}{(2\pi)^{D-1}}
\Biggl(\frac{\q_1^{\,2}\q_2^{~\,'2}+\q_2^{\,2}
\q_1^{~\,'2}}{\vks}-\q^{\,2} \Biggr)~. \label{K B} \ee As for the
infrared divergency, it is convenient, following
Refs.~\cite{Fadin:2006ha} and~\cite{Fadin:2007de}, to regularize
it introducing  the cut-off $\lambda$, tending to zero after
taking the limit $\epsilon\rightarrow0$,  and picking  out in the
representations (\ref{omega s}) and (\ref{omega B}) the domains
$\vec {k}^{\;2}\leq\lambda^{2}$ and
$(\vec{k}-\vec{q}_{i})^{2}\leq\lambda^{2}$. It is easy to see that
the contributions of these domains cancel the  contribution of
$\hat{{\cal K}}^S_n$ from  the region $\vec
{k}^{\;2}\leq\lambda^{2}$. In the remaining regions we can put
$D=4$. Taking into account the transformation  (\ref{trans}) with
$\hat{O}$ given by the relation (\ref{trans-1}) and omitting the
terms leading to $\delta(\xp_1-\xp_2)$, we obtain \[ \langle
\q_1,\q_2|\hat{\omega}_1+\hat{\omega}_2+\hat{{\cal
K}}^S_n|\qp_1,\qp_2\rangle \rightarrow \frac{\alpha_{s}
^{2}(\mu)N_{c}^{2}}{4\pi^{3}}\frac{n_S}{6}\Biggl[-\delta(\q_1-\qp_1)
\delta(\q_2-\qp_2)\int d\vec{k}\left( 2V_S(\vec{k})\right.
\]
\be \left.+V_S(\vec{k},\vec{k}-\vec{q}_{1})+V_S(\vec{k}
,\vec{k}+\vec{q}_{2})\right)+2\delta(\vec{q}-\vec{q}^{\;\prime})\left(
V_S(\vec{k})+V_S(\vec{k},\vec{q} _{1})+V_S(\vec{k},\vec{q}
_{2})\right)\Biggr]~,\label{simplified NA} \ee where \be
V_S(\vec{k})=\frac{1}{2\vec{k}^{\;2}}\left(\ln\left(
\frac{\vec{k}^{\;2}}{\mu^{2}}\right)-\frac{8}{3}\right) ~,
\label{function V1} \ee \be
V_S(\vec{k},\vec{q})=-\frac{\vec{k}\vec{q}}{2\vec{k}^{\;2}\vec{q}^{\;2}}\left(
\ln\left( \frac{\vec{k}^{\;2}\vec{q}^{\;2}}{\mu^{2}(\vec{k}
-\vec{q})^{2}}\right)  -\frac{8}{3}\right) +\frac{1}{4\vec
{k}^{\;2}}\ln\left(
\frac{\vec{q}^{\;2}}{(\vec{k}-\vec{q})^{2}}\right)
\label{function V2} \\
+\frac{1}{4\vec{q}^{\;2}}\ln\left(
\frac{\vec{k}^{\;2}}{(\vec{k}-\vec {q})^{2}}\right)  ~. \ee

The ``Abelian" part can be written in the form
\[
\langle \q_1,\q_2|\hat{{\cal K}}^S_a|\qp_1,\qp_2\rangle =
\delta(\q-\qp)\frac{g^4N_c^2n_S}{4(2\pi
)^{D-1}}\frac{1}{\qs_1\qs_2}\int_0^1 dx\int \frac{d^2k_1}{(2\pi
)^{D-1}} x(1-x)\]
\begin{equation}
\times \left(\frac{\vec{q}_{1}^{\:2}-2(
\vec{q}_{1}\vec{k}_{1})}{\sigma_{11}}+\frac{\vec{q}_{1}^{\:2}-2(\vec{q}_{1}
\vec{k}_{2})}{\sigma_{21}}\right)
 \left( \frac{2(\vec{q}_{2}\vec{k}_{1})+\vec{q}
_{2}^{\:2}}{\sigma_{12}}+\frac{2(\vec{q}_{2}
\vec{k}_{2})+\vec{q}_{2}^{\:2}}{\sigma_{22}}\right) \;, \label{K S
a through f}
\end{equation}
where $\vk_1+\vk_2=\vk=\q_1-\qp_1=\qp_2-\q_2$,
\[
\sigma_{11}=(\vk_1-x\q_1)^2+x(1-x)\qs_1,
\;\;\sigma_{21}=(\vk_2-(1-x)\q_1)^2+x(1-x)\qs_1\;,
\]
\begin{equation} \sigma_{12}=(\vk_1+x\q_2)^2+x(1-x)\qs_2,
\;\;\sigma_{22}=(\vk_2+(1-x)\q_2)^2+x(1-x)\qs_2\;. \label{sigma}
\end{equation}
This part agrees with the forward QED kernel considered in
Ref.~\cite{Kotikov:2000pm}. It contains neither ultraviolet nor
infrared singularities and therefore it does not require neither
regularization nor renormalization. Therefore we can use from the
beginning the physical space-time dimension $D=4$ and the
renormalized coupling constant $\alpha_s$. Restricting ourselves
to the dipole form, we can perform the change
\begin{equation}
\langle \q_1,\q_2|\hat{{\cal K}}^S_a|\qp_1,\qp_2\rangle
\rightarrow  \delta(\q-\qp)\frac{\alpha_s{^2}(\mu)N_c^2n_s}{4\pi
^3}\frac{1}{\qs_1\qs_2}\int_0^1 dx\int \frac{d^2k_1}{(2\pi
)}\frac{x(1-x)}{\sigma_{11}\sigma_{22}}({\vec{q}_{1}^{\:2}-2(
\vec{q}_{1}\vec{k}_{1})(\vec{q}_{2}
\vec{k}_{2})+\vec{q}_{2}^{\:2}})~.\label{simplified K S a} \ee

The transformation of Eqs.~(\ref{simplified NA}) and (\ref{simplified K S
a}) into the dipole form can be easily done with the help of the
integrals presented in Ref.~\cite{Fadin:2007de}. As a result we have
 \be
g_S(\x_1,\x_2;\vrho)=-g^0_S(\x_1,\x_2;\vrho)=
\frac{n_S}{12}\left(\frac{\x_{12}^2}{\x_{1\rho}^2\x_{2\rho}^2}\ln\frac{\x_{S}^{\,2}}
{\x_{12}^2}+\frac{\x_{1\rho}^2-\x_{2\rho}^2}{\x_{1\rho}^2\x_{2\rho}^2}
\ln\frac{\x_{1\rho}^2}{\x_{2\rho}^2}\right)~, \ee where \be\ln
r_{S}^{\,2}=-\frac{8}{3}+2\psi(1)-\ln\frac{\mu^2}{4}~,\label{g3 s}
\ee and \be
g_{S}(\x_1,\x_2;\xp_1,\xp_2)=\frac{n_S}{2}\frac{1}{\x^{\,4}_{1'2'}}
\biggl(\frac{\x^{\,2}_{12'}\x^{\,2}_{1'2}}{d}\ln\frac{\x^{\,2}_{12'}
\x^{\,2}_{1'2}}{\x^{\,2}_{11'}\x^{\,2}_{22'}}-1\biggr)~,\label{g4
s} \ee with $d$ defined in Eq.~(\ref{d}).  At that
$g_S(\x_1,\x_2;\vrho)$ is determined by the ``non-Abelian" part
(\ref{simplified NA}) and $g(\x_1,\x_2;\xp_1,\xp_2)$ by the
``Abelian" part (\ref{simplified K S a}).  It is evidently
conformal invariant.

\section{Discussion}

The results (\ref{g 0 G})-(\ref{g4 M}) and (\ref{g3 s})-(\ref{g4
s}) permit to write the dipole form of the BFKL kernel in
sypersymmetric Yang-Mills theories with any $N$. The most
interesting is the $N=4$ case, with $n_M=4, \;\;n_S=6$, where one
can hope on conformal invariance. Unfortunately, the results
presented in this paper do not show this property. There are terms
violating conformal invariance both in
$g(\vec{r}_{1},\vec{r}_{2};\vec{\rho})$ and in
$g(\vec{r}_{1},\vec{r}_{2};\vec{r}_{1}^{\;\prime},\vec{r}_{2}^{\;\prime})$.
Let us remind, however, that the NLO kernel  is not unambiguously
defined. The transformation (\ref{trans}),  accompanied by the
corresponding transformation of the impact factors, does not
change scattering amplitudes.  The meaning of these
transformations is the redistribution of the radiative corrections
between the kernel and the impact factors. In this aspect  the
situation is quite analogous to the possibility to  redistribute
the radiative corrections between the anomalous dimension and the
coefficient functions.  The transformation (\ref{trans}) can be
used for the elimination of the terms violating conformal
invariance. Till now the possibility of a complete elimination of
these terms is neither proved nor disproved.

Recently the paper~\cite{Balitsky:2007wg} appeared in the web. In
this paper the NLO gluon contribution to the BK kernel is found.
The result of Ref.~\cite{Balitsky:2007wg} is not in  accord with
that of Ref.~\cite{Fadin:2007de}. One could think that the
disagreement can be removed using the transformation
(\ref{trans-1}), but the result of Ref.~\cite{Balitsky:2007wg}
does not agree also with that of Ref.~\cite{FL98} confirmed in
Ref.~\cite{Vogt:2004mw} and Ref.~\cite{Kotikov:2004er}. We hope to
turn to this discrepancy elsewhere.

\vspace{1.0cm} \noindent {\Large \bf Acknowledgment}
\vspace{0.5cm}

V.S.F. thanks the Dipartimento di Fisica dell'Universit\`a della
Calabria and the Istituto Nazionale di Fisica Nucleare, Gruppo
Collegato di Cosenza, for the warm hospitality while part of this
work was done and for the financial support.

\end{document}